\documentclass[12pt]{article}

\usepackage{scicite}

\usepackage{times}

\usepackage[english]{babel}
\usepackage[utf8]{inputenc}
\usepackage[T1]{fontenc}

\usepackage{amsfonts,amssymb,amsmath}
\usepackage[svgnames]{xcolor}
\usepackage{graphicx}
\usepackage[separate-uncertainty=true]{siunitx}
\usepackage[unicode=true,pdfusetitle,bookmarks=true,bookmarksnumbered=true,bookmarksopen=false, breaklinks=false,pdfborder={0 0 0},pdfborderstyle={},backref=false,colorlinks=false]{hyperref}

\DeclareSIUnit\gauss{G}

\hypersetup{pdftitle = {Observing crossover between quantum speed limits},
 pdfauthor = {Gal Ness, Manolo R. Lam, Wolfgang Alt, Dieter Meschede, Yoav Sagi, Andrea Alberti},
	colorlinks=true,linkcolor=blue,citecolor=blue,filecolor=blue,urlcolor=blue,pdfpagemode=UseNone,pdfstartview={XYZ null null 1.00},pdfborder={0 0 0}}

\usepackage{nicefrac,mathtools,cancel,braket,comment,textcomp,xspace,cleveref}

\newcommand{\wHO}{\omega_{\mathrm{HO}}}
\newcommand{\pihalf}{\nicefrac{\pi}{2}}
\newcommand{\lc}{\nicefrac{\lambda}{2}}

\newcommand{\FS}{FS\xspace}
\def\H{\hat{H}}

\newcommand{\figref}[2]{\hyperref[#1]{\ref{#1}{(#2)}}}

\newcommand{\encapsulateMath}[1]{\raisebox{0pt}[0pt][0pt]{#1}}

\makeatletter

\addto\captionsenglish{}
\def\@caption@fignum@sep{{\bfseries .} }
\let\orig@make@capt@title=\@make@capt@title
\def\@make@capt@title#1#2{\orig@make@capt@title{{\bfseries #1}}{#2}}%
\makeatother

\def\supmat#1{see Methods, Sec.~\ref{#1}}

\def\textemdash{\leavevmode\unskip\kern1pt---\kern1pt\ignorespaces}

\newenvironment{sciabstract}{%
\begin{quote} \bf}
{\end{quote}}

\title{Observing crossover between quantum speed limits} 

\author
{Gal Ness,$^{1}$ Manolo R.~Lam,$^{2}$ Wolfgang Alt,$^{2}$ Dieter Meschede,$^{2}$\\Yoav Sagi,$^{1}$ Andrea Alberti$^{2\ast}$\\
\\
\normalsize{$^{1}$Physics Department, Technion -- Israel Institute of Technology, IL-32000 Haifa, Israel}\\
\normalsize{$^{2}$Institut für Angewandte Physik, Universität Bonn, 53115 Bonn, Germany}\\
\\
\normalsize{$^\ast$To whom correspondence should be addressed; E-mail:  alberti@iap.uni-bonn.de.}
}

\date{}

\begin{document}

\baselineskip24pt

\maketitle 

\begin{sciabstract}
Quantum mechanics sets fundamental limits on how fast quantum states can be transformed in time.
Two well-known quantum speed limits are the Mandelstam--Tamm
and the Margolus--Levitin
bounds, which relate the maximum speed of evolution to the system's energy uncertainty and mean energy, respectively.
Here, we test concurrently both limits in a multi-level system by following the motion of a single atom in an optical trap using fast matter wave interferometry.
Our data reveal two different regimes:
one where the Mandelstam--Tamm limit constrains the evolution at all times, and a second where a crossover to the Margolus--Levitin limit is manifested at longer times.
 We take a geometric approach to quantify the deviation from the speed limit,
measuring how much the matter wave's quantum evolution deviates from the geodesic path in the Hilbert space of the multi-level system.
Our results, establishing quantum speed limits beyond the simple two-level system, are important to understand the ultimate performance of quantum computing devices
and related advanced quantum technologies.
\end{sciabstract}

\paragraph*{Introduction}

The celebrated energy-time uncertainty relation was given a rigorous interpretation by Mandelstam and Tamm (MT) as a lower bound on the time it takes a quantum system to evolve into a different state \cite{Mandelstam:1945}.
A second independent bound was formulated by Margolus and Levitin (ML) in terms of the mean energy relative to the ground state \cite{Margolus:1998}.
The maximum of these two times provides a unified bound for the quantum speed limit \cite{Giovannetti:2003a,Levitin:2009,Deffner:2017}.
This limit is relevant to understand the ultimate performance of quantum devices.
It was studied in relation to quantum computing \cite{Lloyd:2000,Lloyd2002,Svozil2005,Santos:2015%
}, parameter estimation in quantum metrology \cite{Giovannetti:2011,Froewis2012}, quantum information transfer \cite{Murphy:2010}, quantum optimal control \cite{Caneva:2009}, and thermodynamic devices such as quantum engines and batteries \cite{Campo:2014,Campaioli:2017}.%

In the simplest scenario of a two-level system (qubit), both MT and ML limits yield the same minimum time to reach an orthogonal state, which is the Rabi flopping time.
The same holds for systems that can be effectively mapped onto two-level Hamiltonians, as was demonstrated by previous experimental investigations \cite{Bason2011,Cimmarusti:2015,Frank:2016,Vepsalainen:2019}.
However, quantum simulation and information processing devices rely on a far greater number of states, and often include a nonvanishing coupling to the continuum.
It is therefore essential to test quantum speed limits beyond the restricted Hilbert space of a qubit.

In this work, we study quantum speed limits in a clean manifestation of a multi-level system \textemdash a single atom in a potential well of finite depth.
The potential supports many bound states, yet at the same time it possesses a continuum of free-particle states, which allow the atom to leave the trap.
Using a fast excitation-interrogation scheme, we investigate the ideal case of a time-independent Hamiltonian $\hat{H}$, where the quantum dynamics originates from an initial motional excitation of the atom \textemdash a matter wave.
In the limit of small excitations, we recover the qubit case, where the quantum evolution  involves mainly two states.
However, by increasing the excitation extent, we depart from this limit in a well-controlled manner to probe the multi-level contribution, up to the point that the atom populates mostly unbound states in the continuum.
Our measurements reveal that both speed limits provide relevant bounds on the system's quantum dynamics.
This result is in stark contrast to the case of a driven multi-level system, where the MT bound yields an excessively short time scale \cite{Lam:2021}.

The MT bound constrains \cite{Fleming:1973,Bhattacharyya:1983,Anandan:1990} the two-time state overlap $\left|\Braket{\psi(0)|\psi(t)}\right|$ from below by means of the energy uncertainty $\Delta E$,
\begin{subequations}\label{eq:QSL}
\begin{equation}\label{eq:MTB}
\left|\Braket{\psi(0)|\psi(t)}\right|\ge\cos\left(\frac{\Delta E t}{\hbar}\right),
\end{equation} 
in the domain $0\le t \le \tau_\text{MT} \equiv \pi\hbar/(2\Delta E$).
Here, $\tau_\text{MT}$ is the MT orthogonalization time, i.e., the minimum duration for the evolved state to become orthogonal to the initial one.
The energy uncertainty follows\cite{fn:bracket_notation} the conventional definition \encapsulateMath{$\Delta E^2 = \braket{\H^2}-\braket{\H}{}^{\hspace{-2pt}2}$}.

The ML bound, on the other hand, constrains\cite{Giovannetti:2003a} the two-time state overlap from below by the mean energy $E=\braket{\hat{H}}$,
\begin{equation}\label{eq:MLB}
\left|\Braket{\psi(0)|\psi(t)}\right|\ge\cos\left(\hspace{-2pt}\sqrt{\frac{\pi E t}{2\hbar}}\right),
\end{equation}
\end{subequations}
in the domain $0\le t \le \tau_\text{ML} \equiv \pi \hbar/(2 E)$, with the ground state energy chosen to be zero. Similarly, $\tau_\text{ML}$ represents the minimum orthogonalization time according to the Margolus-Levitin bound.

The left-hand side of Eqs.\,\eqref{eq:MTB} and \eqref{eq:MLB} can be understood as a measure of the change of the time-evolved quantum state with respect to the original one.
In fact, the two-time state overlap relates directly to the distance covered by the quantum state as measured by the Fubini-Study (FS) metric in the projective Hilbert space \cite{Wootters:1981}, $\mathcal{D}\left[\psi(0),\psi(t)\right] \equiv \arccos\left|\braket{\psi(0)|\psi(t)}\right|$.
This definition of distance allows interpreting the two inequalities as bounds on the quantum state's rate of change, i.e., as quantum speed limits.

\paragraph*{Fast matter wave interferometry}\label{sec:experiment}

The basic idea of our experiment is as follows.
We start with an atom in the vibrational level $n$ of an optical trap. The trap originates from a single site of a one-dimensional optical lattice, with a period of $\lc=\SI{433}{\nano\meter}$  (\supmat{app:experimental}). Because of the deep lattice potential, tunneling between adjacent sites can be completely neglected when the atom populates the low energy states. Subsequently, we suddenly displace the trap minimum by a distance $\Delta x$ and let the atom slide down the potential hill. Finally,  after a time $t$, we use fast matter wave interferometry to measure how far its quantum state has evolved.
To speed up data acquisition, we average over an ensemble of about \num{20} atoms, sparsely filling the one-dimensional lattice.

\begin{figure}
	\centering\includegraphics[width=0.7\textwidth]{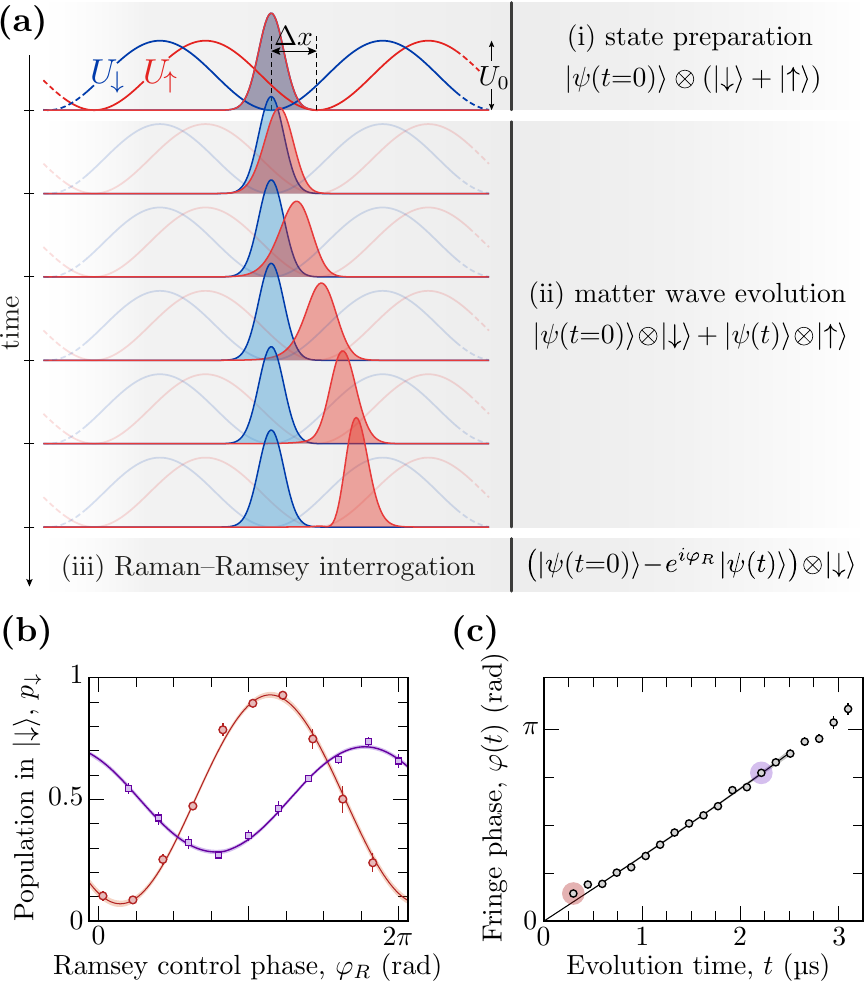}
	\caption{
		\textbf{Fast matter wave interferometry for testing quantum speed limits.}
	{(a)} Illustration of the Raman-Ramsey measurement technique:
	(i) At $t\,{=}\,0$, the atom is placed in a superposition of $\ket{\uparrow}$ and $\ket{\downarrow}$ states, each subject to a different periodic potential, $U_\uparrow$ and $U_\downarrow$.
		(ii) The atom with $\ket{\uparrow}$, initially displaced by $\Delta x$ from the trap center, slides downhill and concurrently deforms by the anharmonicity of its potential.
	The atom with $\ket{\downarrow}$ is in a vibrational eigenstate ($n{=}0$ in this example), which remains unchanged.
	(iii) 
		The probability of occupying $\ket{\downarrow}$ is measured as a function of the control Ramsey phase $\varphi_R$.
					The quantum states are displayed on the right-hand side up to a normalization factor.
		{(b)} Ramsey fringes measured as a function of $\varphi_R$ for two selected evolution times, \SI{300}{\nano\second} and \SI{2.2}{\micro\second},
		with $\Delta x=0.2\,\lc$.
Solid lines are cosine functions fitted to the data, with shades denoting the 1-$\sigma$ confidence regions.
Data points are normalized to account for atom losses ($\SI{5}{\percent}$), and error bars mark the standard error.
{(c)} Fringe phase tracked as a function of time.
Circled points correspond to the fringes displayed in (b).
Solid line is a fifth-order polynomial fit containing only odd-power terms \cite{fn:TIS}, used to extract $E$ based on Eq.~(\ref{eq:phase_expansion}).
\label{fig:experimental}}
\end{figure}

Our interferometry technique is illustrated in Fig.~\figref{fig:experimental}{a}.
At $t\,{=}\,0$, we put the atom in an equal superposition of two internal states, $\ket{\uparrow}$ and $\ket{\downarrow}$, using a fast Raman pulse (\supmat{app:fastRC}).
Each spin state is subject to a different potential \cite{Robens:2018}, $U_\uparrow$ and $U_\downarrow$ (\supmat{app:lattice}).
The atom in state $\ket{\uparrow}$ experiences a $\Delta x$-displaced potential, as described above.
Conversely, the atom in state $\ket{\downarrow}$ is maintained unchanged (up to a global phase) in the vibrational eigenstate $n$ of its trapping potential, where it is originally prepared\cite{Belmechri:2013} before applying the Raman pulse.
Thus, by such a splitting of the matter wave, we effectively create two copies of the same state, where one undergoes the intended downhill evolution, and the other remains stationary, serving as a reference for the state at $t=0$.

After a given evolution time $t$, we let the two copies interfere with each other by applying a second fast Raman pulse, akin to a Ramsey interrogation scheme.
Crucially, both Raman pulses must be much shorter than the time scale for the quantum state evolution, $\max\{\tau_\text{MT},\tau_\text{ML}\}$, which we anticipate to be in the microsecond range.
In the experiment, we achieve pulse durations as short as $\SI{45}{\nano\second}$,
thus ensuring that their action is nearly instantaneous and not affected by the trapping potential.
Also, the two laser beams driving the Raman pulses are co-propagating  (\supmat{app:fastRC}), ensuring that the momentum transferred to the atoms by the Raman process is negligible.

With this fast interrogation technique, we obtain all three quantities needed to test Eqs.~\eqref{eq:MTB} and \eqref{eq:MLB}: the two-time state overlap $\left|\Braket{\psi(0)|\psi(t)}\right|$ as a function of time $t$, the mean energy $E$, and the energy uncertainty $\Delta E$.
To this purpose, we record the probability, $p_\downarrow$, to find the atom in state $\ket{\downarrow}$ as a function of the Ramsey control phase $\varphi_R$, i.e., the relative phase between the first and second pulse.
This measurement yields a typical Ramsey fringe [Fig.~\figref{fig:experimental}{b}],
characterized by a visibility $\mathcal{V}(t)$ and a phase $\varphi(t)$ (\supmat{app:overlapSignature}).
Importantly, these two quantities combined yield the complex-valued overlap integral, $\Braket{\psi(0)|\psi(t)} = \mathcal{V}(t)\exp\{-i[\varphi(t)+E_n t/\hbar]\}$, where
$E_n$ is the energy of the stationary state $n$, with the ground state energy chosen to be zero ($E_0 \equiv 0$). 
Thus, the visibility directly gives us the two-time state overlap, i.e., the first of the three quantities to be measured.

\begin{figure*}[t]
	\centering\includegraphics[width=\textwidth]{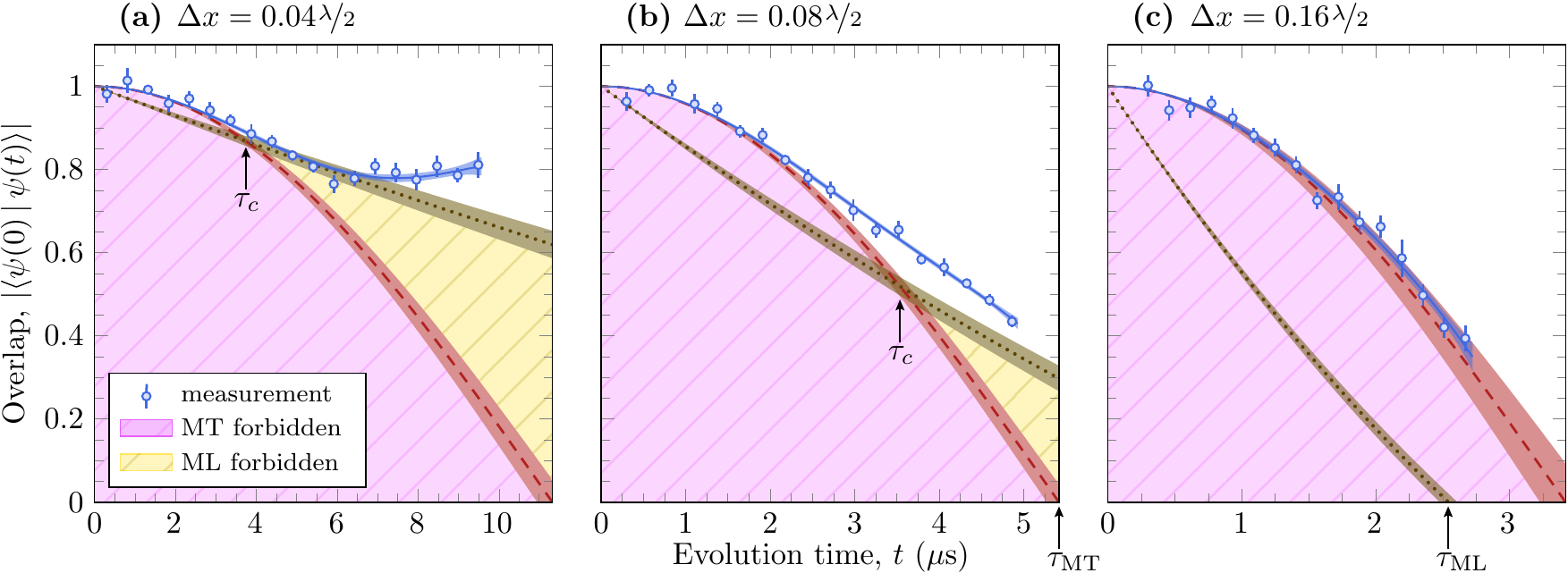}
	\caption{\label{fig:shortTimes}
	\textbf{Quantum speed limits in a multi-level quantum system}.
	Measured two-time state overlap vs evolution time for three displacements $\Delta x$ from the trap center. The initial state is chosen with $n\,{=}\,0$.
		Colored regions are those excluded by the MT (pink) and ML (yellow) bounds.
	The crossover time between the two speed limits is marked by $\tau_c$.
		A sixth-order polynomial containing only even-power terms \cite{fn:TIS} is fitted to the data points (solid line), from which we extract $\Delta E$ using Eq.~(\ref{eq:visibility}).
		Shades around lines represent the 1-$\sigma$ confidence region.
		Note that the $x$-axis domain extends up to $\tau_\text{MT}$, whose value differs in each panel.
							}
\end{figure*}

We obtain $E$, the second quantity to be measured, from the phase of the Ramsey fringe, $\varphi(t)$,
by expanding it for short times \cite{Sjoqvist:2000},
\begin{equation}
	\label{eq:phase_expansion}
	\varphi(t) = \left(E-E_n\right)t/\hbar + O(t^3)\,,
\end{equation}
and knowing the energies $E_n$ from sideband spectroscopy measurements \cite{Belmechri:2013}.
Hence, tracking the phase evolution for short times, we extract $E$ from the linear term of a fifth-order polynomial fit [Fig.~\figref{fig:experimental}{c}].

From the short-time expansion of the visibility \cite{Anandan:1990}, we obtain the third quantity to be determined, $\Delta E$:
\begin{equation}
	\label{eq:visibility}
	\mathcal{V}(t) = 1-(\Delta Et/\hbar)^2/2 + O(t^4)\,.
\end{equation}
This expansion establishes a relation between the short-time evolution and $\Delta E$, which is well recognized in the literature on the quantum Zeno effect \cite{Pascazio:2014}. Thus, we obtain $\Delta E$ by fitting to the measured visibility  a polynomial curve (solid lines in Fig.~\ref{fig:shortTimes}).
It is important to underline that the MT bound of Eq.~(\ref{eq:MTB}) is a statement about the quantum evolution speed that, unlike Eq.~(\ref{eq:visibility}), is not constrained to the short-time limit.

\paragraph*{Results}

In Fig.~\ref{fig:shortTimes} we present three representative data sets of the two-time state overlap, with $n=0$ and initial displacements $\Delta x$ set to (a) $0.04\,\lc$, (b) $0.08\,\lc$, and (c) $0.16\,\lc$.
Comparing the three data sets, we find that the two-state overlap drops at a faster rate for increasing values of $\Delta x$, meaning that the matter wave departs from its original state at a higher speed for increasingly larger excitations.
We compare the data points to the lower bounds as predicted by the MT and ML speed limits in Eqs.~(\ref{eq:MTB}) and (\ref{eq:MLB}), using the values of $\Delta E$ and $E$ extracted from the fitting models in Eqs.~(\ref{eq:visibility}) and (\ref{eq:phase_expansion}).
The regions excluded by the two bounds are hatched in different colors.
The remaining region is the one allowed by the unified bound, defined by the maximum of the two limits.
From this comparison, we make two important observations.
The first is that all data points fall within the allowed region, thus giving the first
experimental confirmation of the unified bound.
Deviations from this bound are quantitatively studied below.
The second observation is that a crossover between the two limits is manifested in panels \hyperref[fig:shortTimes]{(a)} and \hyperref[fig:shortTimes]{(b)}:
The two-time state overlap is bounded from below by the MT bound for short times ($t\,{<}\,\tau_c$) and by the ML for longer times ($t\,{>}\,\tau_c$).

\paragraph*{Quantum-speed-limit crossover}

To research the condition and origin of this crossover, we test a wide spectrum of experimental conditions, leveraging the great degree of control and flexibility of our setup:
The initial displacement can be controlled with sub-nanometer precision over the full range, $0<\Delta x\leq0.5\,\lc$.
We excite mostly bound states for $\Delta x\,{\ll}\, \lc$ and, vice versa, mostly unbound states in the continuum for $\Delta x$ at around $0.5 \,\lc$.
Large displacements, $\Delta x>0.25\,\lc$, allow us, in particular, to test the speed limit for a nonharmonic potential, where the curvature of the potential is inverted.
Furthermore, we vary the type of excitation by choosing the shape of the initial atomic wave packet to have $n=\{0,1,2\}$ nodes along the direction of the motional excitation.
Since states with $n>0$ differ starkly from Gaussian-like states, their quantum evolution is substantially different from that of semiclassical matter waves.

\begin{figure}
	\centering\includegraphics[width=0.7\textwidth]{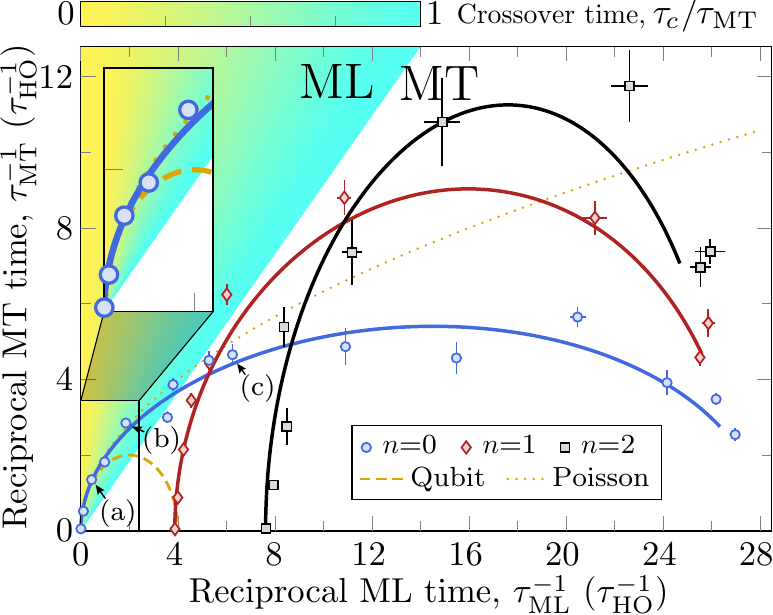}
	\caption{\textbf{Quantum-speed-limit crossover}.
		Measured orthogonalization times, $\tau_\text{ML}$ and $\tau_\text{MT}$, displayed through their reciprocals, with $n$ the quantum number characterizing the initial wave packet shape.
	Shades in color identify the ML regime, where a crossover manifests at time $\tau_c$, as opposed to the MT regime, where no crossover occurs.
		Inset highlights data points in the ML regime.
	The three points marked with arrows correspond to panels (a), (b), and (c) of Fig.~\ref{fig:shortTimes}.
					Solid lines show the expected curves computed with no free fitting parameters by numerical diagonalization of $\hat{H}$ (\supmat{app:lattice}).
			The limiting case of a qubit (dashed line) and a coherent excitation (dotted line) are also shown.
	Values are expressed in units of the reciprocal of the trap oscillation period, which is around $\SI{16}{\micro\second}$.
						\label{fig:EvsDE}}
\end{figure}

We examine $\num{34}$ combinations of parameters, and record for each of them a data set as those shown in Fig.~\ref{fig:shortTimes}.
Inspecting each data set individually shows that the vast majority are bounded by the MT limit only.
However, in a few cases, a crossover to the ML bound is manifested at longer times, as exemplified in panels \hyperref[fig:shortTimes]{(a)} and \hyperref[fig:shortTimes]{(b)}.
To explain the crossover condition, we display in Fig.~\ref{fig:EvsDE} the extracted energy uncertainty $\Delta E$ and mean energy $E$, in terms of the reciprocal of the MT and ML orthogonalization times.
The inset highlights cases where the crossover is clearly visible.
Considering the functional form of Eqs.\,\eqref{eq:MTB} and \eqref{eq:MLB}, we find that the crossover occurs when the orthogonalization times satisfy the condition $\tau_\text{MT}<\tau_\text{ML}$.
The region defined by this condition is identified in the diagram by shades of color, with the color representing the crossover time, $\tau_c = \tau_\text{MT}^2/\tau_\text{ML}$ ($0<\tau_c<\tau_\text{MT}$).
This expression defines a general time scale, which applies to any quantum system evolving under a static Hamiltonian, for it derives directly from Eqs.\,\eqref{eq:MTB} and \eqref{eq:MLB}.
When $\tau_\text{MT}<\tau_\text{ML}$, we call it the ML regime, since the quantum state evolution is constrained by the ML bound for $t>\tau_c$.
Conversely, we call $\tau_\text{MT}>\tau_\text{ML}$ the MT regime, since the evolution is solely constrained by the MT bound for all times.

To obtain insight into the origin of the crossover, we gather the points in three different groups according to the quantum number $n$.
Remarkably, the points falling in ML region are only those with $n=0$, and in particular those in the limit $\tau_\text{MT}^{-1} \lesssim \tau_\text{HO}^{-1}$, where $\tau_\text{HO}\equiv 2\pi/\wHO$ is the trap oscillation period in the harmonic approximation.
This limit corresponds to small initial excitations, $\Delta E \lesssim \hbar\wHO$, when
only very few levels are involved: mainly the original vibrational level $n$ and with a small probability the additional levels $n\pm 1$ (or only level \num{1} when $n\,{=}\,0$).
To understand why this limit falls into the ML regime, it is sufficient to consider the limiting case of a qubit subject to a static Hamiltonian.
Representing the qubit as a spin precessing around a fixed axis at frequency $\wHO/(2\pi)$  (\supmat{app:effectiveQubit}), one finds that the condition for the ML regime, $\tau_\text{ML}>\tau_\text{MT}$, is fulfilled as long as the lower level is more populated than the upper level.

By contrast, for large excitations, the distribution of the many excited levels is highly localized as a function of energy ($\Delta E < E$), yielding an evolution in the MT regime ($\tau_\text{MT} > \tau_\text{ML}$).
An example interpolating the two limiting cases of small and large excitations is obtained by considering a coherent excitation in a harmonic potential (dotted line), 
where the population of the vibrational levels follows a Poisson distribution.
Notably, this curve fits well only the $n=0$ series for sufficiently small excitations.
Its failure to fit the rest of the data reveals that the matter waves tested here include but are not limited to the semiclassical case of coherent excitations.

\paragraph*{Deviation from the speed limit}\label{sec:deviation}

To gain insight into the mechanisms leading to deviations from the speed limit, we specifically consider the MT bound because it applies to both regimes.
For a quantitative analysis, we take a geometric point of view, as proposed by Anandan and Aharonov \cite{Anandan:1990}, which relies on the \FS metric as a measure of the distance between states in Hilbert space.
They showed that the length of the path traced by the time-evolved state equals $\ell(t) \equiv \pi t/(2\tau_\text{MT})$.
On the other hand, the length of the shortest path (geodesic) connecting the initial state to that at time $t$ amounts to the FS distance between the two states, \encapsulateMath{$\ell_\text{geo}(t) \equiv \mathcal{D}\left[\psi(0),\psi(t)\right]$}.
The MT bound in Eq.~(\ref{eq:MTB}) can be expressed as $\ell(t) \geq \ell_\text{geo}(t)$, which has a clear geometrical interpretation \textemdash
the MT bound is saturated only when the system evolves along a geodesic \cite{Uhlmann:2009}.
Using the definition of $\mathcal{D}$ and the expansion in Eq.~(\ref{eq:visibility}), we express the geodesic length as a series of powers of $t/\tau_\text{MT}$:
\begin{equation}
\label{eq:FS_expansion}
\ell_\text{geo}(t)/\ell(t) = 1- \frac{\pi^2 \xi}{48} \left(\frac{t}{\tau_\text{MT}}\right)^2  + O(t^4)\,,
\end{equation}
where $\xi$ is a dimensionless parameter empirically introduced to quantify the deviation of the state evolution from the geodesic.
The foregoing geometrical condition, representing the MT bound, translates into $\xi \geq 0$.

\begin{figure}[t]
	\centering\includegraphics[width=0.7\textwidth]{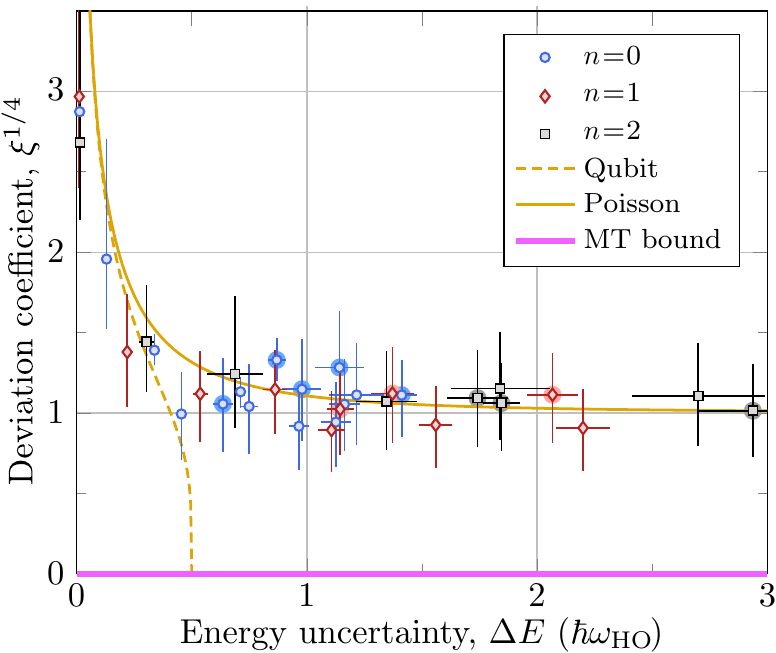}
	\caption{\textbf{Deviations from the MT speed limit.}
		The measured coefficient $\xi$ is plotted vs $\Delta E$ for a wide spectrum of experimental conditions, varying the initial displacement $\Delta x$ and the wave packet shape $n=\{0,1,2\}$.
	The pink line indicates the MT bound.
  Points corresponding to $\Delta x \,{>}\, 0.25\,\lc$ (nonharmonic regime) are highlighted by a surrounding circle.
	Note that for better visualization purposes the fourth root of $\xi$ is plotted, as $\xi$ is related to the fourth moment of the spectral distribution.
							\label{fig:deviations}}
\end{figure}

We obtain the deviation coefficient $\xi$ from a
polynomial fit of the measured \FS distance $\ell_\text{geo}$ for each data set examined above.
The results are displayed in Fig.~\ref{fig:deviations} as a function of the energy uncertainty $\Delta E$.
As expected, all data points fall within the allowed region.
Surprisingly, however, the points coalesce around $\xi=1$
regardless of the wave packet shape, the potential's anharmonicity, and for nearly all values of $\Delta E$, save for the limiting case of very small excitations, discussed below.
Such a coalescence hides a nontrivial relation with the energy uncertainty.
In fact, for this result to hold, $t$ in the power series in Eq.~(\ref{eq:FS_expansion}) must be expressed in units of $\tau_\text{MT}$, which in turn depends on $\Delta E$.

We attribute the observed strict deviations from the MT bound, $\xi>0$, to the multi-level nature of our system.
This result is in line with the well-known fact that only a qubit system can evolve along a geodesic \cite{Uhlmann:2009}, and thus, saturate the MT bound.
For a quantitative interpretation of the deviation coefficient, we derive its expression in terms of $\hat{H}$ from the unitary evolution underlying the Schrödinger equation, $\xi=(\beta_2-1)/2$, where $\beta_2 = \langle{(\hat{H}-E)^4\rangle}/\Delta E^4$ is the kurtosis of the energy spectrum.
This expression reduces to $\xi=1$ if we model the population of the vibrational levels by a Gaussian distribution as a function of energy.

It is nonetheless remarkable that the observed deviations are small on the scale of $\tau_\text{MT}$.
We explain this observation by the well-known fact in statistics that many tailed distributions relevant to describe energy excitations, such as coherent excitations, have a kurtosis around \num{3}, thus yielding $\xi \approx 1$ (\supmat{app:positiveXi}).
Owing to the small factor $\pi^2\xi/48 $ in Eq.~(\ref{eq:FS_expansion}),
we therefore conclude
that the MT speed limit establishes a relevant bound on the evolution rate of a multi-level system subject to a time-independent Hamiltonian [see, e.g., Fig.~\figref{fig:shortTimes}{c}], in clear contrast to what observed in time-driven multi-level systems \cite{Lam:2021}.

The limit of small excitations, $\Delta E \lesssim \hbar \wHO$, reveals qualitatively different physics, with values of $\xi$ significantly larger than \num{1}.
To obtain further insight, we consider the limiting case of a qubit, for which
we find $\xi_\text{qubit}=2(\Delta E_\text{max}^2/\Delta E^2-1)$, where $\Delta E_\text{max}=\hbar\wHO/2$ is the maximum energy uncertainty attainable by the qubit.
This expression reveals a good agreement with the experimental data.
It also shows that the bound is theoretically saturated ($\xi_\text{qubit}=0$) for $\Delta E=\Delta E_\text{max}$, which occurs when the spin precesses along a great circle (the geodesic).
In practice, however, this situation never occurs in our setup, since small excitations of the matter wave correspond to the case of the spin forming a very small angle with respect to the precession axis.
In this limit, the evolved state is far from becoming orthogonal to the original one and, correspondingly, deviations from the MT bound are large on the scale of the orthogonalization time $\tau_\text{MT}$.
This situation is well exemplified by Fig.~\figref{fig:shortTimes}{a}, which shows a cosine-like oscillation of the two-time state overlap, corresponding to a spin precessing with a small angle of about $\SI{40}{\degree}$.
An interpolation between the qubit case ($\xi\gg1$) and the foregoing multi-level Gaussian case ($\xi=1$) is obtained considering the previous example of a Poisson distribution, which yields
$\xi_\text{HO}=1+(\hbar\wHO)^2/(2\Delta E^2)$.
The comparison of this curve with the experimental data shows an excellent agreement,
which also holds for $n=\{1,2\}$ (\supmat{app:deviation_coef}).

\paragraph*{Discussion}

Our study sheds light on two fundamental limits of quantum dynamics.
We have tested them concurrently using matter wave interferometry in an infinite-dimensional system.
The observation of a crossover between the two limits shows indeed that the unified bound is what constrains the quantum dynamics of a system's excitation.
Such a crossover bears implications for any quantum system governed by a static Hamiltonian, for it is expected to occur whenever $\tau_\text{ML}>\tau_\text{MT}$, i.e., when the energy uncertainty is larger than the mean energy.
This is particularly for the case of small excitations, when only a few levels are involved, as shown by our measurements probing the dynamics of a matter wave.
Through our measurements, we have uncovered the relevance of multiple levels in determining the evolution rate.
Excitations of a multi-level system do not saturate the speed limit but, unexpectedly, reveal a universal deviation from it.
Importantly, this deviation is found to be small, despite the well-known fact that the MT bound can only be attained in a qubit system.
Its small magnitude is explained as a consequence of the small kurtosis of the spectrum of the matter wave excitations.
We expect similar behavior to apply to a large number of systems when excitations have a short-tailed spectrum, establishing the MT bound as a useful figure for the evolution time scale.
Notable exceptions however exist, when the excitation has a heavy-tailed spectrum, as is the case of the Breit-Wigner distribution \cite{Sakurai:1994}.
{} 

Key to our study is the ability to measure the two-time state overlap, which gives the FS distance covered by the evolving state.
We emphasize the direct nature of this measurement, which, leveraging matter wave interferometry, does not require quantum state tomography nor any prior knowledge of the spectrum of $\hat{H}$.
This technique is reminiscent of that used in Loschmidt echo experiments \cite{Andersen:2003,Cetina:2016}, with the notable difference that one of the two branches of the Ramsey interferometer is held in our case stationary, so as to measure the FS distance from the initial state.
Also, we remark that our measurement of the FS distance is not limited to the manifold of states reachable by an adiabatic transformation, where the metric reduces to that defined by the geometric quantum tensor \cite{Gianfrate:2020}.

This work deals with quantum dynamics on the time scale of $\tau_\text{MT}$.
However, the same matter wave interferometer technique developed here can be used to explore quantum state evolution on a much longer time scale \cite{Lam:2021a}.
Furthermore, the same technique opens up the possibility to directly measure the FS distance in time-dependent systems, relevant for testing the quantum speed limit in the presence of an external drive \cite{Deffner:2013b,Okuyama:2018a}.
We also envisage extensions of this technique to open quantum systems in order to measure the Bures distance\textemdash Uhlmann's generalization \cite{Uhlmann:1992} of the FS distance to mixed states \textemdash for unitary \cite{Mondal:2016} or non-unitary \cite{Funo:2019} evolutions.
Understanding the quantum speed limit of open quantum systems is an important ongoing effort \cite{Taddei:2013,Campo:2013,Zhang:2014,Pires:2016,Campaioli:2018,Deffner:2020,OConnor:2021}.

In the future, our results may find applications in quantum systems involving multiple levels, such as atomtronics \cite{Amico:2020}, bosonic quantum computing \cite{Blais:2021},
and quantum simulations in multi-level systems \cite{Chen:2021}.

\section*{Methods}

\def\thesubsection{\Alph{subsection}}%
\def\paragraph*#1{\subsection{#1}}%

\subsection{Experimental sequence}\label{app:experimental}

An ensemble of $^{133}$Cs atoms is cooled in a three-dimensional magneto-optical trap (MOT) and subsequently transferred into an optical trap consisting of a one-dimensional optical lattice formed by two counter-propagating laser beams with wavelength $\lambda \approx \SI{866}{\nano\meter}$.
A small ensemble of about twenty atoms is sparsely loaded into the one-dimensional optical lattice, with a vanishing probability of having more than one atom per lattice site due to losses induced by light-assisted collisions.
The initial number of atoms is measured by collecting the fluorescence light emitted by the atoms when these are illuminated by nearly resonant laser beams.
Subsequently, the laser beams are kept on for an additional \SI{10}{\milli\second} with a reduced intensity and a larger detuning in order to cool the atoms into a low energy motional state.
During loading, detection, and cooling of atoms, the lattice depth $U_0$ is set sufficiently large ($k_B\times\SI{370}{\micro \kelvin}$) to suppress the probability that an atom hops between lattice sites. Here, $k_B$ denotes the Boltzmann constant.

For the preparation of the atom state and the subsequent experiments testing the quantum speed limits, the lattice depth is reduced to about $k_B \times  \SI{26}{\micro \kelvin}$, corresponding to $\num{270}\,E_R$, where $E_R = (2\pi\hbar)^2/(2m\lambda^2)$ is the recoil energy of an atom of mass $m$.
Owing to the large value of $U_0/E_R$, tunneling between sites is completely negligible (\supmat{app:lattice}) when the atoms occupy a low energy motional state.
For the low energy motional states, the trap potential can be approximated by a harmonic oscillator, with trap frequencies $\omega_\text{HO} \approx 2\pi\times\SI{66}{\kilo\hertz}$ in the direction longitudinal to the lattice, and  $\approx 2\pi\times\SI{1}{\kilo\hertz}$ in the transverse directions.
We note that the harmonic approximation in the longitudinal direction only applies in the limit of small excitations, i.e., for $\Delta x \lesssim \sqrt{\hbar/(m \omega_\text{HO})}$, or equivalently, $\Delta E \lesssim \hbar\omega_\text{HO}$.
Also, due to the large difference between the two trap frequencies, the excitations in the longitudinal direction and the transverse directions are decoupled.
Such a decoupling between the longitudinal and transverse motional degrees of freedom
allows us to treat the quantum evolution of the matter wave as an effectively one-dimensional problem.

We employ microwave sideband cooling \cite{Belmechri:2013} to cool the atoms along the longitudinal direction into the vibrational ground state and, simultaneously, optically pump them to the Zeeman state $\ket{F=4,m_F=4}$ of the electronic ground state.
A bias magnetic field of $\SI{3}{\gauss}$ oriented in the lattice direction is used to define the quantization axis.
The ground state population of the longitudinal motion is measured to be around $\gtrsim \SI{96}{\percent}$ using sideband spectroscopy.
Subsequently, a microwave $\pi$ pulse transfers the atoms to state $\ket{F=3,m_F=3}$.
By tuning the pulse frequency to be resonant with one of the motional sidebands, we selectively transfer the atoms into the desired vibrational level $n$ of the $U_\downarrow$ potential.
The pulse fidelities are $\SI{95}{\percent}$, $\SI{85}{\percent}$ and $\SI{68}{\percent}$ for the eigenstates $n=\{0,1,2\}$, respectively.
The atoms that are not successfully transferred remain in $\ket{F=4,m_F=4}$, and removed from the trap using an optical push-out pulse.

With the atom initialized in the vibrational level $n$, we adiabatically vary (\supmat{app:lattice}) in \SI{300}{\micro\second} the relative position of the two lattices to reach the desired displacement, $0<\Delta x \leq 0.5\,\lc$, and then carry out the matter wave interferometer sequence (\supmat{app:overlapSignature}), which consists of two fast $\pi/2$ pulses separated by a time $t$ and resonant with the transition between the internal states $\ket{\downarrow} = \ket{F=3,m_F=3}$ and $\ket{\uparrow}=\ket{F=4,m_F=3}$.

After the second $\pi/2$ pulse, we remove the atoms in $\ket{\uparrow}$ with a second optical push-out pulse, increase the lattice depth, illuminate the atoms with nearly resonant light, and collect the emitted fluorescence light. 
We re-normalize the detected fluorescence by the fidelity of the $\pi$ pulse used to prepare the atom in the vibrational level $n$, in order to compensate for the fraction of atoms removed by the first push-out pulse.
The ratio between the re-normalized fluorescence and the initially detected one yields an estimate of $p_\downarrow$.
To gain sufficient statistics, in addition to averaging over the ensemble of atoms loaded into the one-dimensional lattice, we average over \num{10} repetitions of the sequence described above.

\subsection{Fast Raman pulse setup}\label{app:fastRC}

We employ a pair of phase-locked laser beams to drive the fast pulses of the matter wave interferometer by means of resonant two-photon transitions.

Before illuminating the atoms, the two beams are coupled into a common optical fiber and then sent through a double-pass acousto-optic modulator (AOM).
By controlling the RF drive power of the AOM, we can temporally shape the intensity of the Raman pulses with nanosecond precision.
On such a short time scale, the AOM intensity control exhibits a nonlinear response, which is taken into account and compensated using a look-up table.
A second optical fiber is employed to overlap the Raman laser beams with one of the two laser beams forming the one-dimensional optical lattice (\supmat{app:experimental}).
Thereby, we ensure that the Raman beams overlap perfectly with the optical trap, and that the momentum transferred to the atom by the Raman transition is negligible, since the two Raman beams are co-propagating.

The two Raman beams are red-detuned by about $2\pi\times\SI{48}{\giga\hertz}$ from the cesium $\text{D}_2$ line.
Because of the large detuning, the probability of an off-resonant photon scattering event during the pulse duration is negligible ($\approx \num{e-4}$).
The frequency difference of the Raman beams is tuned to the hyperfine splitting of about $2\pi\times\SI{9.2}{\giga\hertz}$, separating the two internal states $\ket{\uparrow}$ and $\ket{\downarrow}$.
The two Raman laser beams illuminate the atoms with an individual power of $\SI{1.2}{\milli\watt}$ and the same circular polarization.
This polarization enables $\sigma$-type transitions, since the quantization axis is collinear with the optical lattice (and, thus, with the Raman beams).

We achieve a high effective Rabi frequency $\Omega_R \approx 2\pi\times \SI{6.5}{\mega\hertz}$ as a result of the high intensity of the Raman beams, which are tightly focused onto the atoms through a relatively small waist ($\approx \SI{17}{\micro \meter}$).
Such a high intensity causes, in addition, a differential light shift of about $\SI{12}{\mega\hertz}$, which is taken into account by tuning the frequency difference of the two Raman beams to be resonant with the shifted transition.
In the Ramsey interrogation scheme, this differential light field adds to the phase of the Ramsey fringe a linear shift in time by \SI{81}{\radian\per\micro\second}, which is subsequently subtracted from the Ramsey phase $\varphi(t)$.

The Rabi frequency $\Omega_R$ is mostly homogeneous over the entire ensemble of atoms, thus ensuring a high fringe visibility $\gtrsim \SI{96}{\percent}$.
We observe relative variations of less than one percent for atoms positioned at different lattice sites due to the collinearity of the Raman and lattice beams.
The atoms have a small, but not vanishing temperature of about $\SI{1.5}{\micro \kelvin}$ in the directions transverse to the lattice, resulting in a distribution of the atoms' transverse positions and thus of the Rabi frequency.
To reduce the inhomogeneous spread of the Rabi frequency, we use an additional blue-detuned hollow beam counter-propagating to the Raman beams to increase the confinement of atoms close to the optical axis, where the intensity of the Raman laser beams is maximal.

\subsection{Spin-dependent optical lattice setup}
\label{app:lattice}

The spin-dependent optical lattice setup is described in detail in Ref.~\cite{Robens:2018}, where a schematic illustration is also provided.
We employ two counter-propagating laser beams, each linearly polarized with an angle $\theta$ between their linear polarizations, to create two superimposed optical standing waves of right- and left-handed circularly polarized light.
Then, by controlling $\theta$, we displace the two standing waves along their common axis by $\Delta x_\text{sw}(\theta) = (\theta/\pi)\lc$ with sub-nanometer precision.

The light shift exerted on the atoms by the two standing waves gives rise to two spin-dependent optical lattices, $U_\uparrow$ and $U_\downarrow$, which differ for the two spin states because of their specific polarization-dependent ac polarizability.
At the wavelength $\lambda$, one can show that the potential $U_\uparrow$ comprises two contributions proportional to the intensity of the right circularly polarized light (relative weight $\num{7/8}$) and the intensity of the left circularly polarized light (relative weight $\num{1/8}$);
the same expression holds for $U_\downarrow$, with the two polarization circularities being exchanged.

The two potentials, $U_\uparrow$ and $U_\downarrow$, exhibit an ideal sinusoidal profile along the longitudinal direction, with a lattice constant equal to $\lc$.
Due to both standing waves contributing to the lattice potential, the displacement $\Delta x$ between the two lattices has a slightly nonlinear dependence on the polarization angle,
\begin{equation}
	\Delta x(\theta) = \frac{\tan^{-1}[\nicefrac{3}{4}\,\tan(\theta)]}{\theta}  \Delta x_\text{sw}(\theta)\,.
\end{equation}
The trap depth $U_0$ is equal for both lattices.
However, it slightly depends on the polarization angle $\theta$,
\begin{equation}
		U_0(\theta)/U_0(0) =  \sqrt{\frac{25+7\cos(2\theta)}{32}} \end{equation}
again as a result of the contribution by both standing waves.
The minimum trap depth is $U_0(\nicefrac{\pi}{2}) = \nicefrac{3}{4}\,U_0(0) \approx 200\,E_R$ occurring when $\Delta x(\nicefrac{\pi}{2})=0.5\,\lc$.
Because the trap depth varies with $\theta$ (equivalently, $\Delta x$), the trap frequency also slightly depends on the displacement, $\omega_\text{HO}(\theta)=\sqrt{2U_0(\theta)/(m \lambda^2)}$.

In the deep lattice regime, $U_0\gg E_R$, the atom has a negligible probability to tunnel to adjacent sites when it populates a low energy state.
The low energy bands are virtually flat, and correspondingly the tunneling time is much longer than the microsecond time scale of the experiment.
For example, the band with index $n=0$ has a tunneling time greater than $\num{1}$ year.
By contrast, the energy bands with $n\gtrsim U_0/(\hbar \omega_\text{HO}) \approx 10 $ resemble the dispersion relation of a free particle.
The tunneling time varies nearly exponentially as a function of the band index $n$, changing by more than \num{12} orders of magnitude from $n=0$ to $n=10$.
Hence, it is a very good approximation to consider states belonging to the low energy bands as effectively bound states, and vice versa states in the higher energy bands as unbound states.

Because states in the low energy bands have negligible tunneling, it suffices to consider the lattice Hamiltonian on a single site, assuming periodic boundary conditions, in order to numerically compute the initial wave packet for $n=\{0,1,2\}$.
Thus, the Hamiltonian of the single lattice site is evaluated on a discrete grid of \num{1024} spatial points, and used to calculate the low-lying eigenstates by numerical diagonalization. The eigenstates so obtained are then extended by zero padding, shifted by $\Delta x$, and then used to compute the mean energy $E$ and energy uncertainty $\Delta E$ of the optical lattice Hamiltonian $\hat{\mathcal{H}}$ over a lattice consisting of several sites.
The theoretical curves in Fig.~\ref{fig:EvsDE} showing $\tau_\text{MT}^{-1}$ as a function of $\tau_\text{ML}^{-1}$ are directly obtained from the computed values of $\Delta E$ and $E$.

\subsection{Matter wave interferometry}\label{app:overlapSignature}

The matter wave interferometer sequence shown in Fig.~\figref{fig:experimental}{a} is  described in detail below.
At time $t=0$, the atom occupies $\ket{\psi(0)} \otimes \ket{\downarrow}$, where $\ket{\psi(0)}\equiv \ket{n}$ is one of the motional eigenstates of $U_\downarrow$ potential.
The first $\pihalf$ pulse, acting nearly instantaneously (\supmat{app:fastRC}), puts the atom in a superposition both of spin states,
\begin{equation}
\frac{1}{\sqrt{2}}\ket{\psi(0)} \otimes (\ket{\downarrow} + \ket{\uparrow})\,,
\end{equation}
where the atom's motional state remains unchanged during the short pulse.
Afterwards, the atom is let evolve for a duration $t$, resulting in
\begin{equation}
\frac{1}{\sqrt{2}}\left[e^{-i E_n t/\hbar}\ket{\psi(0)} \otimes \ket{\downarrow} + \ket{\psi(t)} \otimes \ket{\uparrow}\right],
\end{equation}
expressed in the frame rotating with the hyperfine frequency of the transition between the two spin states.
A second fast $\pihalf$ lets the two branches of the superposition state interfere with each other, yielding
\begin{equation}
\frac{e^{-i E_n t/\hbar}\ket{\psi(0)}- e^{i\varphi_R}\ket{\psi(t)} }{2}\otimes \ket{\downarrow} + 
\frac{e^{-i (E_n t/\hbar +\varphi_R)}\ket{\psi(0)}+\ket{\psi(t)} }{2}\otimes \ket{\uparrow},
\end{equation}
where $\varphi_R$ is the Ramsey control phase, which is varied by controlling the relative phase between the first and second pulse.
Finally, a push-out pulse removes the atoms in state $\ket{\uparrow}$, and the probability $p_\downarrow$ of occupying state $\ket{\downarrow}$ is measured as a function of $\varphi_R$, producing a typical Ramsey fringe,
\begin{equation}
	p_\downarrow(\varphi_R) = \frac{1 - \mathcal{V}(t) \cos[\varphi_R -\varphi(t)]}{2}\,,
\end{equation}
where the visibility $\mathcal{V}(t)$ and phase $\varphi(t)$ are related to the complex-valued overlap integral, $\Braket{\psi(0)|\psi(t)} = \mathcal{V}(t)\exp\{-i[\varphi(t)+E_n t/\hbar]\}$.
The fringe phase is shifted by an offset, $E_n t/\hbar$, where $E_n$ is the energy of the vibrational level $n$ with respect to the ground state, $E_0=0$, known by sideband spectroscopy \cite{Belmechri:2013}.

\subsection{Tailedness of spectral distribution as a measure of deviation from MT bound}
\label{app:positiveXi}

In the main text, the MT bound is shown to imply the inequality $\xi\geq 0$,
where $\xi$ is a coefficient accounting for the tailedness (kurtosis) of the spectral distribution of the excitation,
\begin{equation}
	\label{eq:xi}
	\xi = \frac{\langle{(\hat{H}-E)^4\rangle}-\langle(\hat{H}-E)^2\rangle^2}{2\langle(\hat{H}-E)^2\rangle^2}\,.
\end{equation}
A normal distribution has a deviation coefficient $\xi=1$.

The distribution is said to be leptokurtic for $\xi>1$ and platykurtic for $\xi<1$.
Leptokurtic are most of the tailed distributions describing excitations in a many-level system.
By contrast, the most platykurtic distribution is notably the Bernoulli distribution with an equal probability of heads and tails, for which the deviation coefficient reaches its minimum possible value, $\xi=0$.
Such a distribution describes the excitation of a qubit with both eigenstates equally populated.
Since the MT bound is saturated for this excitation, we thereby prove that $\xi=0$ is not only a necessary but also a sufficient condition to saturate the MT bound.

It is interesting to observe that the numerator in Eq.~(\ref{eq:xi}) is equal to the variance of \encapsulateMath{$(\hat{H}-E)^2$}, and must therefore be positive, thus providing an independent confirmation of the result derived from the MT bound.

An upper bound for the deviation coefficient $\xi$ can be obtained when the spectral distribution of the excitation is bounded from above by an energy cutoff $E_c$.
In this case, by making use of the Bhatia--Davis inequality \cite{Bhatia:2000}, which constrains the variance of a random variate with a bounded probability distribution, we obtain the bound:
\begin{equation}
    \label{eq:BhatiaDavis}
    \xi\le\frac{1}{2}\left[\max\left\{ \left(\frac{E_{c}-E}{\Delta E}\right)^{2},\left(\frac{E}{\Delta E}\right)^{2}\right\} -1\right],
\end{equation}
where we used the fact observed above that the variance of $(\hat{H}-E)^2$ is equal to the numerator of Eq.~(\ref{eq:xi}).
From the properties of the Bhatia--Davis inequality, one finds that the upper bound in Eq.~(\ref{eq:BhatiaDavis}) is saturated when the excited states have their energy concentrated at either one of the endpoints of the spectrum.
This occurs, in particular, in the limit of small excitations, referred to in the main text as the qubit case (see also Methods, Sec.~\ref{app:effectiveQubit}), for which the ground state is populated with probability close to one.
The right-hand side of Eq.~(\ref{eq:BhatiaDavis}) reproduces, in fact, the exact behavior of $\xi_\text{qubit}$ presented in the main text, diverging as $\hbar\omega_\text{HO}/(2\Delta E)$ for vanishingly small excitations.
It should, however, be added that the bound in Eq.~(\ref{eq:BhatiaDavis}) is far from tight in the case of many relevant spectral distributions, such as the spectrum of a truncated coherent excitation.
In this case, the spectrum resembles that of a Gaussian distribution when $E\ll E_c$, producing $\xi\approx 1$.
In contrast, the right-hand side of Eq.~(\ref{eq:BhatiaDavis}) tends to $E/(2\hbar\omega_\text{HO})$, which is much larger than \num{1} when $\hbar \omega_\text{HO}\ll E$.

\subsection{Deviation coefficient in the harmonic approximation}
\label{app:deviation_coef}

At $t=0$, the excited motional state is equal to the vibrational eigenstate $\ket{n}$ displaced by $\Delta x$,
\begin{equation}
\label{eq:excited_state}
\ket{\psi(0)} = e^{-i \hat{p} \Delta x/\hbar} \ket{n},
\end{equation}
where $\hat{p}$ is the momentum operator.
In the harmonic approximation, valid for sufficiently small excitations, the probability distribution for the case $n=0$ is given by
\begin{subequations}
\begin{equation}
	p_{n=0}(n')=\left|\Braket{n' \hspace{-1pt}| \psi(0) }\right|^2 = \frac{e^{-|\alpha|^2}}{n'!} \left| \alpha \right|^{2n'}\!,
\end{equation}
where $|\alpha| = \sqrt{m\,\wHO/(2\hbar)}\, \Delta x$ is the amplitude of the corresponding coherent state ($m$ is the atomic mass).
Using $p_{n=0}(n')$ to compute $\xi$ in Eq.~(\ref{eq:xi}), one obtains
$\xi_{n=0}=1+(\hbar\wHO)^2/(2\Delta E^2)$.
Note that in the main text $\xi_{n=0}$ is denoted as $\xi_\text{HO}$.

For the other cases, $n=1$ and $n=2$, the probability distributions in the harmonic approximation are
{\makeatletter
\@centering=-25pt plus 1pt%
\makeatother
\begin{eqnarray}
p_{n=1}(n')&=&\frac{\left(\left|\alpha\right|^2 -n' \right)^2}{|\alpha|^2}\,p_{n=0}(n')\,,\\
p_{n=2}(n')&=&\frac{\left(\left|\alpha\right|^4-2r\left|\alpha\right|^2 +n'^2-n'\right)^2}{2|\alpha|^4}\,p_{n=0}(n')\,,\end{eqnarray}
}%
\end{subequations}
yielding the coefficients $\xi_{n=1}=1/3+(\hbar\wHO)^2/(2\Delta E)^2$ and $\xi_{n=2}=7/25+(\hbar\wHO)^2/(2\Delta E)^2$, respectively.
Notably, the expression of $\xi$ exhibits for all three cases the same behavior in the limit of very small excitations, where mainly two states ($n=0$) and three states ($n>0$) are excited.

\subsection{The qubit case}
\label{app:effectiveQubit}

For very small excitations, $|\alpha| \ll 1$, the excited state in Eq.~(\ref{eq:excited_state}) in the case of $n=0$ reduces to two levels,
\begin{equation}
	\label{eq:qubit_approx}
	\ket{\psi(0)} \approx  \ket{0} + |\alpha|\, \ket{1},
\end{equation}
as in a qubit system.

In general, a qubit precessing with frequency $\omega_\text{HO}/(2\pi)$ around a fixed axis at an angle $\zeta$ has $\Delta E = \hbar \wHO \sin(\zeta)/2 $, $E=\hbar\wHO \sin^2(\zeta/2)$
and the two-time state overlap
\begin{equation}
	\label{eq:qubit_overlap}
	\left|\Braket{\psi(0)|\psi(t)}\right|  = \sqrt{1-\sin^2(\zeta)\sin^2(\wHO t/2)}\,.
\end{equation}
In the case the two states are equally populated, $\zeta=\pi/2$, the two-time state overlap in Eq.~(\ref{eq:qubit_overlap}) saturates the MT bound in Eq.~(\ref{eq:MTB}) for all times, $0\leq t\leq \tau_\text{MT}$.
In the case of no population inversion ($0<\zeta<\pi/2$), as in Eq.~(\ref{eq:qubit_approx}), we have $\Delta E>E$, meaning that the qubit is in the ML regime.
For the other case of population inversion ($\pi/2<\zeta<\pi$), we remark that a bound equivalent to the ML bound in Eq.~(\ref{eq:MLB}) can be derived considering the fact that the energy is limited from above,
\begin{equation}
	\left|\Braket{\psi(0)|\psi(t)}\right| \ge \cos \hspace{-1pt}\left[\hspace{-1pt}\sqrt{ \cos^2(\zeta/2)\pi \wHO t/2}\, \right].
\end{equation}

To compute the deviation coefficient $\xi_\text{qubit}$, which is discussed in the main text, we directly apply Eq.~(\ref{eq:xi}) to a Bernoulli distribution with $p_0 = \cos(\zeta/2)^2$ and $p_1=1-p_0$.

\section*{Acknowledgments}
The authors would like to thank Jan Uckert for his contribution to the setup, Antonio Negretti for his critical reading, and Nir Davidson for helpful discussions.
Also, the authors wish to pay tribute to Leonid Mandelstam for his pioneering contribution, together with G.~Landsberg, to the discovery \cite{Landsberg:1928} of Raman transitions, which are key to the fast interrogation scheme of this work.
We thank C.~Feuersänger for developing PGFPlots, which was used in this work.
\textbf{Funding:} This research was supported by the Reinhard Frank Foundation in collaboration with the German Technion Society, and the SFB/TR 185 OSCAR of the German Research Foundation. 
G.~N.~is supported by the Helen Diller Quantum Center at the Technion, and M.~R.~L. by the German Academic Exchange Service.
\textbf{Author contributions:} A.A.\ and Y.S.\ conceived the experiment.
G.N.\ and M.R.L.\ built the Raman setup and conducted the measurements.
G.N., M.R.L., and A.A.\ analyzed the data.
A.A.\ and G.N.\ wrote the manuscript with inputs from all authors.
All authors contributed to the discussions and analysis of the results.
\textbf{Competing interests:} The authors declare that they have no competing interests.
\textbf{Data and materials availability:} All data needed to evaluate the conclusions in the paper
are present in the paper.

\section*{Supplementary materials}
Methods


\begin{thebibliography}{10}
\bibitem{fn:bracket_notation}
The angle bracket notation denotes the expectation value of
  \encapsulateMath{$\hat{H}$}, or powers thereof, computed for the state
  \encapsulateMath{$\ket{\psi(t)}$}; as \encapsulateMath{$\hat{H}$} is time
  independent, so is its expectation value.
\bibitem{fn:TIS}
The even-power terms in Eq.~\eqref{eq:phase_expansion}
  and the odd-power terms in Eq.~\eqref{eq:visibility} vanish due to the
  time-inversion symmetry of $\hat{H}$. Both fitting models for
  the visibility and the phase use three fitting parameters, thus neglecting
  corrections of order $\mathcal{O}(t^7)$. The order of the fitting polynomial
  is chosen not too large to avoid over-fitting..
\end{thebibliography}
\end{document}